\documentclass[showpacs,preprintnumbers,amsmath,amssymb,prd]{revtex4}

\begin{document}
\title{Symmetry breaking in nonuniform noncommutative $\lambda\phi^4$ theory at finite temperature }
\author{J. M. Hern\'andez}
\email{javierh@fcfm.buap.mx}
\affiliation{Benem\'erita Universidad Aut\'onoma de Puebla, Facultad de Ciencias F\'{\i}sico Matem\'aticas, P.O. Box 165, 72000 Puebla, M\'exico.}
\author{C. Ram\'{\i}rez}
\email{cramirez@fcfm.buap.mx}
\affiliation{Benem\'erita Universidad Aut\'onoma de Puebla, Facultad de Ciencias F\'{\i}sico Matem\'aticas, P.O. Box 165, 72000 Puebla, M\'exico.}
\author{M. S\'anchez}
\email{hanamichi2_2000@yahoo.com}
\affiliation{Benem\'erita Universidad Aut\'onoma de Puebla, Facultad de Ciencias F\'{\i}sico Matem\'aticas, P.O. Box 165, 72000 Puebla, M\'exico.}

\begin{abstract}
We consider the 2PI Cornwall-Jackiw-Tomboulis effective action at finite temperature for a noncommutative real scalar field theory  in 4 dimensions, with noncommutativity among space and time variables. By means of a Rayleig-Ritz variation, we study the solutions of a stripe type nonuniform background, which depends on space and time, and hence on temperature. The analysis in the first approximation shows that such solutions appear in the planar limit, as already known, but also under normal noncommutativity, in an anisotropic region which has not been considered. Further we show that the transition from the uniform ordered phase to the non uniform one is first order.
\end{abstract}
\maketitle

\section{Introduction}
Noncommutativity in field theory has received much attention in recent years, as a means to explore short distances, i.e. high energy effects \cite{szabo}. In particular these effects could be due to a fundamental theory behind, like string theory \cite{sw}. An interesting point which has attracted attention, is the possibility to observe low energy noncommutative effects due to the UV/IR mixing \cite{chen}. In this direction, an interesting question is the finite temperature behavior and phase transitions. In particular there is a breaking of translational invariance \cite{gubser}, which leads to the appearance of a stripe phase, related to the UV/IR mixing \cite{minwalla,gubser,ambjorn}. In this paper it was considered a phase diagram with a Lifshitz point which was also studied in \cite{chen}. In these works it is argued that there is a phase transition from the uniform to the non uniform ordered phases, although it is not clear what happens at four dimensions. 
Moreover, in the study of noncommutative theories, the noncommutativity of time with space variables is known to lead to causality problems \cite{seiberg}. However, at finite temperature, in the imaginary time formalism, there is no reason why not to consider full noncommutativity \cite{strelchenko}. In fact, the noncommutativity of the imaginary time could be related to noncommutative effects regarding temperature, suggesting a relation with nonequilibrium processes. Such processes have been studied in \cite{calzetta,berges0} by means of the effective action of Cornwall, Jackiw and Tomboulis (CJT) \cite{cjt}, which is given by an expansion in two-particle irreducible (2PI) Feynman diagrams. This effective action was proposed for the study of composite particles, and allows also to make selective summations to higher loop graphs \cite{jac}, which on the other side can be seen as consistent truncations which describe dissipative processes \cite{calzetta}. In \cite{castorina1}, the authors argue that in a noncommutative $\lambda\phi^4$ field theory \cite{micu} with a constant background, there is no symmetry breaking of the internal symmetry. However they show that in the stripe phase with space nonuniformity \cite{gubser,brazovski}, symmetry can be broken. In this paper we consider the conditions for the existence of a stripe phase, in a scalar field theory with $\lambda\phi^4$ interaction at finite temperature, with noncommutativity among all variables. Consequently, we include a dependence of the stripe background on the imaginary time, and hence on the temperature. The study non uniform solutions is made by means of a Rayleigh-Ritz variation of the CJT effective action, and we look for consistent stripe solutions. It turns out that there are such solutions in the planar limit, as already found in \cite{chen,gubser}, but also with anisotropic noncomutativity, which was not considered in previous works, with a temperature dependent background. Moreover there is a discontinuity from the uniform phase to the non uniform one, which points to a first order transition between these phases.
The paper is organized as follows, in the second section we review some features of the CJT action, in the third section we consider the noncommutative, finite temperature formulation, in the third section we show the main results of the paper, and in the last section we draw some conclusions.

\section{CJT effective action}

Let us consider an action of scalar fields $I(\Phi)$. The CJT \cite{cjt} effective action $\Gamma(\phi,G)$ is obtained including a quadratic source term in the generating functional of the Green functions,
\begin{equation}
Z(J,K)=\int D\Phi \exp\left\{i\left[I(\Phi)+\int d^4x\Phi(x)J(x)+\frac{1}{2}\int d^4xd^4y\Phi(x)K(x,y)\Phi(y)\right]\right\},
\end{equation}
and then doing a double Legendre transformation of the generating functional of the connected diagrams $iW(J,K)=\log Z(J,K)$. Thus
\begin{eqnarray}
\Gamma(\phi,G)=W(J,K)&-&\int d^4x \phi(x)J(x)-\frac{1}{2}\int d^4xd^4y\phi(x)K(x,y)\phi(y)\nonumber\\
&-&\frac{1}{2}\int d^4xd^4yG(x,y)K(y,x).\label{legendre}
\end{eqnarray}
Defining
\begin{eqnarray}
\frac{\delta W(J,K)}{\delta J(x)}&=&\phi(x),\\
\frac{\delta W(J,K)}{\delta K(x,y)}&=&\frac{1}{2}\left[\phi(x)\phi(y)+G(x,y)\right],
\end{eqnarray}
it turns out that $\phi\left(  x\right)$ is the vacuum expectation value of the field $\Phi(x)$, and $G\left(  x,y\right)  $ is the connected two point function. 
The physical solutions are obtained from the stationarity conditions,
\begin{eqnarray}
\frac{\delta}{\delta\phi(x)}\Gamma\left(  \phi,G\right)&=&0,\\
\frac{\delta}{\delta G(x,y)}\Gamma\left(  \phi,G\right)&=&0.
\end{eqnarray}
The resulting effective action is \cite{cjt}
\begin{equation}
\Gamma\left(  \phi,G\right)  =I\left(  \phi\right)  +\frac{i}{2}Tr\ln
G^{-1}+\frac{i}{2}Tr\left[  \Delta^{-1}\left(  \phi\right)  G\right]
+\Gamma_{2}\left(  \phi,G\right)  -\frac{i}{2}Tr\left(  1\right)  \label{cjt},
\end{equation}
where $\Gamma_{2}\left(  \phi,G\right) $ is the 2PI diagrams expansion,
\begin{equation}
iD^{-1}\left(  x-y\right)  =-\left(  \partial^{\mu}\partial_{\mu}%
+m^{2}\right)  \delta^{4}\left(  x-y\right)  \label{propagador}
\end{equation}
and
\begin{equation}
i\Delta^{-1}\left(  x-y\right)  =\frac{\delta^2 I(\phi)}{\delta\phi(x)\delta\phi(y)}=-\left(  \partial^{\mu}\partial_{\mu}%
+m^{2}\right)  \delta^{4}\left(  x-y\right)  +\frac{\delta^{2}I_{\text{int}%
}\left(  \phi\right)  }{\delta\phi\left(  x\right)  \delta\phi\left(
y\right)  }.\label{delta}
\end{equation}

For a $\lambda \Phi^4$ action
\begin{equation}
I\left(  \Phi\right)  =\int_{-\infty}^{\infty}d^{4}x\left(  \frac{1}%
{2}\partial^{\mu}\Phi\partial_{\mu}\Phi-\frac{1}{2}m^{2}\Phi^{2}-\frac
{\lambda}{4!}\Phi^4\right)  \label{accion0},
\end{equation}
$\Gamma_2$ is given to first order in $\lambda$ by \cite{cjt},
\begin{equation}
\Gamma_{2}\left(  \phi,G\right)  =-\frac{\lambda}{8}\int_{-\infty}^{\infty
}d^{4}xG^{2}\left(  x,x\right)  \label{gama2}.
\end{equation}

\section{Noncommutative finite temperature action}
Let us consider the noncommutative action for a real scalar field with a $\lambda \Phi^4$ potential,
\begin{equation}
I\left(  \Phi\right)  =\int_{-\infty}^{\infty}d^{4}x\left(  \frac{1}%
{2}\partial^{\mu}\Phi\partial_{\mu}\Phi-\frac{1}{2}m^{2}\Phi^{2}-\frac
{\lambda}{4!}\Phi\ast\Phi\ast\Phi\ast\Phi\right)  \label{accion}.
\end{equation}
The noncommutative Weyl-Moyal product is given by,
\begin{equation}
\Phi\left(  x\right)  \ast\Psi\left(  x\right)   =\left.  e^{\frac{i}%
{2}\theta^{\mu\nu}\partial_{\mu}^{x}\partial_{\nu}^{y}}\Phi\left(
x\right)  \Psi\left(  y\right)  \right\vert _{y=x}\label{moyal}.
\end{equation}

We are interested on symmetry breaking of noncommutative finite temperature field theory. Action (\ref{accion}), as well as (\ref{accion0}), is symmetric under $\Phi\rightarrow -\Phi$. The imaginary time formulation is obtained from the relativistic field theory by setting the fields periodic in the imaginary time variable $\tau=i t$, $\Phi(\tau+\beta)=\Phi(\tau)$. The imaginary time integration interval is $[0,\beta]$, and $i\int d^4x\rightarrow\int_0^\beta d\tau\int d^3x$. The fourier transform of the fields is then given by
\begin{equation}
\phi(x)=\frac{1}{\beta}\sum_{-\infty}^{\infty}\int \frac{d^3p}{(2\pi)^3}e^{i(\omega_n\tau+\vec{p}\vec{x})}\phi_n(p)\label{fourier},
\end{equation}
where $\omega_n=\frac{2\pi n}{\beta}$ is the Matsubara frequency, see e.g. \cite{kapusta}. Thus the free action entering into the CJT effective action will be,
\begin{eqnarray}
I_0(\phi)&=&-\frac{1}{2}\int_0^\beta d\tau\int d^3x\left[(\partial_\tau\phi)^2+(\partial_i\phi)^2+m^2\phi^2\right]\nonumber\\
&=&-\frac{1}{2\beta}\sum_{n=-\infty}^{\infty}\int \frac{d^3p}{(2\pi)^3}\left(p_n^2+m^2\right)\phi_n(\vec p)\phi_{-n}(-\vec p)\label{accionnc},
\end{eqnarray}
where $p_n\equiv (\omega_n,\vec p)$, i.e. $p_n^2=\omega_n^2+{\vec p}^2$.

Further, we write the Weyl-Moyal product as
\begin{equation}
\phi\left(  x\right)  \ast\phi\left(  x\right)   
=\left.  
e^{\frac{i}{2}\partial_{x}\wedge\partial_{y}}\phi\left(x\right)  \phi\left(  y\right)  \right\vert _{y=x}
\equiv\left.  
e^{\frac{i}{2}[\Theta_{\tau i}(\partial_{\tau}^{x}\partial_{i}^{y}-\partial_{i}^{x}\partial_{\tau}^{y})+
\Theta_{ij}\partial_{i}^{x}\partial_{j}^{y}]}\phi\left(x\right)  \phi\left(  y\right)  \right\vert _{y=x}
,\label{moyalt}
\end{equation}
where $\Theta_{\tau i}=i\theta^{0i}$ and $\Theta_{ij}=\theta^{ij}$.
Therefore, for the interaction action we get
\begin{equation}
I_{int}(\phi)=\frac{\lambda}{4!}(2\pi)^2\prod_{k=1}^4\left[\sum_{n_k}\int\frac{d^3p}{(2\pi)^3}\phi_{n_k}(\vec{p}_k)\right]\,\,
e^{\frac{i}{2}(p_{n1}\wedge p_{n2}+p_{n3}\wedge p_{n4})}\delta_{\sum_i n_i,0}\,\delta^3\left(\sum_i\vec p_i\right),
\label{accionint}
\end{equation}
where 
\begin{equation}\label{wedge}
p_{n1}\wedge p_{n2}=\Theta_{\tau i}(\omega_{n_1}p_{2i}-\omega_{n_2}p_{1i})+
\Theta_{ij}p_{1i}p_{2j}=\vec\Theta_\tau(\omega_{n_1}\vec p_2-\omega_{n_2}\vec p_1)+\vec\Theta(\vec{p_1}\times\vec{p_2}),
\end{equation}
$(\Theta_\tau)_i=\Theta_{\tau i}$ and $\Theta_i=\epsilon_{ijk}\Theta_{jk}$.

The noncommutative CJT action can be obtained from (\ref{accion}) and (\ref{cjt}), as follows,
\begin{equation}
i\Gamma\left(\phi,G\right)=iI\left(  \phi\right)  -\frac{1}{2}Tr\ln
G^{-1}+\frac{1}{2}Tr\left\{\frac{\delta^2}{\delta\phi\delta\phi} [iI(\phi)]  G\right\}
+i\Gamma_{2}\left(  \phi,G\right)  +\frac{1}{2}Tr\left(  1\right)  \label{cjt1}.
\end{equation}
All terms in this expression, modulo additive constants, become real by the continuation to the imaginary time, and we get, including the nonplanar two loop terms,
\begin{eqnarray}
&&i\Gamma\left(\phi,G\right)=-\frac{1}{2\beta}\sum_l\int \frac{d^3p}{(2\pi)^3}\left(p_l^2+m^2\right)\phi_l(\vec p)\phi_{-l}(-\vec p)\nonumber\\
&+&\frac{\lambda}{4!}\left(\frac{2\pi}{\beta}\right)^3\prod_{k=1}^4\left[\sum_{n_k}\int\frac{d^3p_k}{(2\pi)^3}\phi_{n_k}(\vec{p}_k)\right]\,\,
e^{\frac{i}{2}(p_{n1}\wedge p_{n2}+p_{n3}\wedge p_{n4})}\delta_{\sum_i n_i,0}\,\delta^3\left(\sum_i\vec p_i\right)\nonumber\\
&+&\frac{1}{2}\sum_l\int d^3p\left\{(p_l^2+m^2)G(p_l)+\log[G(p_l)]\right\}\delta^3(\vec 0)\\
&+&\frac{\lambda}{4\beta}\sum_{k,l}\int\frac{d^3p}{(2\pi)^3}\frac{d^3q}{(2\pi)^3}\left[\frac{1}{\beta}\phi_k(\vec p)\phi_{-k}(-\vec p)G(q_l)
-\frac{(2\pi)^3}{2}G(p_k)G(q_l)\delta^3(\vec 0)\right]\left[1+\frac{1}{2}e^{i(p_k\wedge q_l)}\right]+{\rm const.}\nonumber\label{cjt2}
\end{eqnarray}

\section{Nonuniform solutions}
In order to look for nonuniform ground state fields, we will consider such a solution. In \cite{gubser,castorina1}, following \cite{brazovski}, the field is chosen to be given in a stripe phase, $\phi(x)=A\cos(\vec Q\vec x)$, which does not depend on time in order to keep energy conserved. However, considering the possibility of non closed systems \cite{chou}, we will take a dependence on the imaginary time as follows
\begin{equation}
\phi(\tau,\vec x)=A\cos(\omega_n\tau+\vec Q\vec x),\label{stripe}
\end{equation}
which depends on temperature through the Matsubara frequency $\omega_n$. Due to its dependence on only one mode, these fields are commutative, i.e. $\phi(\tau,\vec x)\ast\phi(\tau,\vec x)=\phi^2(\tau,\vec x)$.
Further, for simplicity, following \cite{castorina1} we keep $G(x,y)=G(x-y)$ translational invariant.
The Fourier expansion coefficients of  (\ref{stripe}) are
\begin{equation}
\phi_l(\vec p)=(2\pi)^3\frac{\beta}{2}A\left[\delta_{ln}\delta^3(\vec p-\vec Q)+\delta_{l,-n}\delta^3(\vec p+\vec Q)\right]\label{stripef}.
\end{equation}
The integration of the terms in (\ref{cjt2}) which contain these coefficients, give results discontinuous in  $\vec Q=\vec 0$. For instance
\begin{equation}
-\frac{1}{2\beta}\sum_l\int \frac{d^3p}{(2\pi)^3}\left(p_l^2+m^2\right)\phi_l(\vec p)\phi_{-l}(-\vec p)=-\frac{\beta}{4}A^2\left(Q_n^2+m^2\right)V_3-\frac{\beta}{4}A^2m^2\delta_{n,0}\delta_{\vec Q}V_3,\label{disc}
\end{equation}
where $\delta_{\vec Q}$ vanishes if $\vec Q\neq\vec0$, and is $1$ when $\vec Q=\vec0$, and $V_3=\int d^3x$ is the three dimensional volume. Thus, up to the volume, for $Q_n\neq0$ (\ref{disc}) gives $\frac{\beta}{4}A^2\left(Q_n^2+m^2\right)$, whose limit when $Q_n\rightarrow0$, is $\frac{\beta}{4}A^2 m^2$. However for an uniform background, i.e. $\phi$ constant, (\ref{disc}) gives $\frac{\beta}{2}A^2 m^2$. Note that this discontinuity is the same for a time independent stripe background.

Thus considering these contributions, we get
\begin{eqnarray}
&&i\Gamma\left(\phi,G\right)=\bigg\{-\frac{1}{4}\beta A^2\left(Q_n^2+m^2+\frac{\lambda}{16}A^2\right)\nonumber\\
&+&\frac{1}{2}\sum_l\int \frac{d^3q}{(2\pi)^3}\left\{(q_l^2+m^2)G(q_l)+\log[G(q_l)]\right\}\nonumber\\
&+&\frac{\lambda}{8}A^2\sum_l\int \frac{d^3q}{(2\pi)^3} G(q_l)\left[1+\frac{1}{2}\cos\left(Q_n\wedge q_l\right)\right]\nonumber\\
&-&\frac{\lambda}{8\beta}\sum_{k,l}\int \frac{d^3p}{(2\pi)^3}\frac{d^3q}{(2\pi)^3} G(p_k)G(q_l)\left[1+\frac{1}{2}\cos\left(p_k\wedge q_l\right)\right]\bigg\}V_3\nonumber\\
&+&\frac{1}{4}\beta A^2\left[-\frac{5\lambda}{48}A^2-m^2+\frac{3\lambda}{4\beta}\sum_l\int\frac{d^3q}{(2\pi)^3}G(q_l)\right]\delta_{n,0}\delta_{\vec Q}V_3
+{\rm const}\label{cjt3}.
\end{eqnarray}
Due to the dependence of (\ref{stripe}) on parameters, a Rayleigh-Ritz variation can be performed \cite{cjt}. Thus variating the effective action with respect to $A$, $\vec Q$ and $G(p_k)$, we get respectively the equations
\begin{eqnarray}
A\Bigg\{\frac{\lambda}{8}A^2+Q_n^2+m^2-\frac{\lambda}{2\beta}\sum_l\int \frac{d^3q}{(2\pi)^3} G(q_m)\left[1+\frac{1}{2}\cos(Q_n\wedge q_l)\right]\nonumber\\
+\left[\frac{5\lambda}{24}A^2+m^2-\frac{3\lambda}{4\beta}\sum_l\int\frac{d^3q}{(2\pi)^3}G(q_l)\right]\delta_{n,0}\delta_{\vec Q}
\Bigg\}=0,\label{eqa}\\
A^2\left[-Q_i+\frac{\lambda}{2\beta}\sum_l\int\frac{d^3q}{(2\pi)^3}G(q_l)\left(2\pi l\Theta_{\tau i}T+\epsilon_{ijk}\Theta_jq_k\right)\sin(Q_n\wedge q_l)
\right]=0\label{eqq}
\end{eqnarray}
and
\begin{eqnarray}
&&G^{-1}(p_k)+p_k^2+m^2+\frac{\lambda}{4}A^2\left[1+\frac{1}{2}\cos(Q_n\wedge p_k)\right]\nonumber\\
&&-\frac{\lambda}{2\beta}\sum_l\int \frac{d^3q}{(2\pi)^3}G(q_l)\left[1+\frac{1}{2}\cos(p_k\wedge q_l)\right]+\frac{3\lambda}{8}A^2\delta_{n,0}\delta_{\vec Q}=0\label{eqg}.
\end{eqnarray}
From the third equation, we see that the integrals $\int d^3qG(q_l)$ have UV divergences and in order to regularize them, a high-momentum $\Lambda$ cutoff can be made \cite{gubser}. Further we define the (cut-off regularized) integrals
\begin{eqnarray}
I_1(p_k)&=&-\sum_l\int_\Lambda \frac{d^3q}{(2\pi)^3}\ G(q_l)\cos(p_k\wedge q_l),\label{i1}\\
I(p_k)&=&I_1(0)+\frac{1}{2}I_1(p_k).\label{i}
\end{eqnarray}
A first observation is the discontinuity of equations (\ref{eqa}) and (\ref{eqg}) at $\vec Q=\vec0$, which points to a phase transition from the uniform, constant $\phi$ field, to the stripe phase with $\vec Q\neq\vec0$ and $n\neq0$. Indeed, (\ref{eqa}) has solution $A=0$ and, depending on the parameters, it can have non vanishing solutions which depend on $\vec Q$ as follows,

\noindent For $\phi=$constant, i.e. $\vec Q=\vec0$, $n=0$, defining $A_0=A|_{\phi={\rm constant}}$ and $G_0(q_l)\equiv G(q_l)|_{\phi={\rm constant}}$, we get
\begin{equation}
\frac{\lambda}{6}A_0^2=-m^2+\frac{3\lambda}{4\beta}\sum_l\int_\Lambda \frac{d^3q}{(2\pi)^3}\ G_0(q_l)\label{a0},
\end{equation}
Further for $\vec Q\neq\vec0$, $n\neq0$
\begin{equation}
\frac{\lambda}{8}A^2(Q_n)=-m^2-Q_n^2-\frac{\lambda}{2\beta}I(Q_n)\label{aq},
\end{equation}
whose limit for $\vec Q\rightarrow\vec0$, $n=0$, is
\begin{equation}
\frac{\lambda}{8}A^2(0)=-m^2+\frac{3\lambda}{4\beta}\sum_l\int_\Lambda \frac{d^3q}{(2\pi)^3}\ G(q_l,0)\label{aq0},
\end{equation}
where $G(q_l,0)=G(q_l,Q_n)|_{\vec Q=\vec 0,n=0}$, where we made explicit the dependence on $Q_n$. Clearly (\ref{aq0}) differs from (\ref{a0}), by the factors on the l.h.s.. These equations have solutions, at least for low temperatures, only if $m^2<0$. Thus in the following we will suppose that this is the case and we set $\mu^2=-m^2$.

Further, from (\ref{eqg}) we get for $A=0$, that $G^{-1}(p_k)$ does not depend on $Q_n$,
\begin{equation}
G^{-1}(p_k)=-p_k^2+\mu^2-\frac{\lambda}{2(2\pi)^3\beta}\left.I(p_k)\right|_{A=0}.\label{g0}
\end{equation}
However, if $A\neq 0$, $G^{-1}(p_k)$ depends on $A$, which must be taken from (\ref{a0})-(\ref{aq0}), and we have the following cases.

\noindent For $\phi=$constant
\begin{equation}
G_0^{-1}(p_k)=-p_k^2-\frac{7}{2}\mu^2-\frac{\lambda}{4\beta}\sum_l\int_\Lambda \frac{d^3q}{(2\pi)^3}\ G_0(q_l)\left[\frac{23}{2}-\cos(p_k\wedge q_l)\right]
\label{g0}
\end{equation}
For $\vec Q\neq\vec0$, $n\neq0$, 
\begin{eqnarray}
G^{-1}(p_k,Q_n)&=&-p_k^2+\mu^2+2\left(-\mu^2+Q_n^2\right)\left[1+\frac{1}{2}\cos(Q_n\wedge p_k)\right]\nonumber\\
&&-\frac{\lambda}{2\beta}\left\{\left[1+\frac{1}{2}\cos(Q_n\wedge p_k)\right]I(Q_n)-\frac{1}{2}I(p_k)\right\},
\label{g1}
\end{eqnarray}
from which we get for $\vec Q\rightarrow\vec0$, $n=0$,
\begin{eqnarray}
G^{-1}(p_k,0)&=&-p_k^2-2\mu^2-\frac{\lambda}{4\beta}\sum_l\int_\Lambda \frac{d^3q}{(2\pi)^3}\ G(q_l,0)\left[7-\cos(p_k\wedge q_l)\right].
\label{gq0}
\end{eqnarray}

In order to understand what happens, we need to know under which conditions $\vec Q$ has non vanishing solutions. As (\ref{eqq}) is a very complicated equation, we will look for solutions which, by consistency, are necessarily non vanishing. Obviously, (\ref{eqq}) is identically fulfilled when $A=0$. Otherwise we have
\begin{equation}
Q_i=\frac{\lambda}{2\beta}\sum_l\int_\Lambda\frac{d^3q}{(2\pi)^3}G(q_l)\left(2\pi l\Theta_{\tau i}T+\epsilon_{ijk}\Theta_jq_k\right)\sin(Q_n\wedge q_l).
\label{eqq1}
\end{equation}
In order to analyze it, we define the following adimensional quantities, $\vec\theta=\vec\Theta\Lambda^2$, $\vec\theta_\tau=\vec\Theta_\tau\Lambda^2$, $\vec\kappa=\frac{\vec Q}{\Lambda}$, $\hat\mu=\frac{\mu}{\Lambda}$, $\vartheta=\frac{T}{\Lambda}$ and $r_i=\frac{q_i}{\Lambda}$ for the integration variables, which now run in the interval $[-1,1]$. With these variables (\ref{eqq1}) turns into
\begin{equation}
\kappa_i=-\frac{\lambda \vartheta}{2}\sum_l\int_\Lambda\frac{d^3r}{(2\pi)^3}\frac{\left(2\pi l\vartheta\theta_{\tau i} +\epsilon_{ijk}\theta_jr_k\right)\sin\left[2\pi\vartheta\vec\theta_\tau(l\vec\kappa-n\vec r)+\vec\theta(\vec r\times\vec\kappa)\right]}{{r}^2-\hat\mu^2+\left[2(\hat\mu^2-\kappa^2)-\frac{\lambda \vartheta}{\Lambda}I(\kappa_n)\right]\left\{1+\frac{1}{2}\cos\left[2\pi\vartheta\vec\theta_\tau(l\vec\kappa-n\vec r)+\vec\theta(\vec r\times\vec\kappa)\right]\right\}-\frac{\lambda \vartheta}{2\Lambda}I(r_l)}
\label{kappa}.
\end{equation}
Obviously $\kappa_i$ vanish in the commutative limit. As we are interested on non vanishing solutions, we will look at the limit of the r.h.s., when $\kappa_i\rightarrow 0$. If the integral tends to zero, then $\kappa_i=0$ is a consistent solution, otherwise, if the integral remains finite, then $\kappa_i\neq0$. 

As a first step, we will consider finite values of $\vec\theta_\tau$ and $\vec\theta$, at least one of them non vanishing. A first case is when $\vec\theta_{\tau}=\vec0$ or $n=0$, in this case the integral vanishes as $\kappa_i$ tend to zero, and $\kappa_i=0$ is a consistent solution. On the other side, for non vanishing $\vec\theta_\tau$ and $n$, when $\kappa_i$ tend to zero we get
\begin{equation}
\kappa_i\simeq\frac{\lambda \vartheta}{2}\sum_l\int_\Lambda\frac{d^3r}{(2\pi)^3}\frac{\epsilon_{ijk}\theta_jr_k\sin\left(2\pi n\vartheta\vec\theta_\tau\vec r\right)}{{r}^2+\hat\mu^2\left[1+\cos\left(2\pi n\vartheta\vec\theta_\tau \vec r \right)\right]}+{\cal O}[(\lambda\vartheta)^2]\label{kappan},
\end{equation}
where the term proportional to $2\pi l\vartheta$ in the numerator of (\ref{kappa}) vanishes because the integrand is an odd function. If we decompose now the $\theta_j$ in the numerator of (\ref{kappa}) parallel and perpendicular to $\vec\theta_\tau$, $\vec\theta=\vec\theta_\|+\vec\theta_\perp$, then the part of the integral with $\vec\theta_\|$ will vanish, also because the integrand is odd, as can be easily seen, for instance if $\vec\theta_\tau$ is in the $z$-direction. Thus
\begin{equation}
\kappa_i\simeq\frac{\lambda \vartheta}{2}\sum_l\int_\Lambda\frac{d^3r}{(2\pi)^3}\frac{\epsilon_{ijk}(\theta_\perp)_jr_k\sin\left(2\pi n\vartheta\vec\theta_\tau\vec r\right)}{{r}^2+\hat\mu^2\left[1+\cos\left(2\pi n\vartheta\vec\theta_\tau \vec r \right)\right]}\label{kappan1},
\end{equation}
which does not vanish, as can be verified by a numerical evaluation. Note that the maximal rank parametrization taken in \cite{gubser} for $\Theta_{\mu\nu}$, corresponds to $\vec\theta=\vec\theta_\tau$. Therefore for finite noncommutative parameters, the solutions for $\kappa_i$ of (\ref{kappa}) are non vanishing, if $n$ and $\vec\theta\times\vec\theta_\tau\sim(\Theta^2)_{0i}$ do not vanish. 

As a next step, we consider the planar limit, i.e. $\vec\theta_\tau$ and $\vec\theta$ tend to infinity, and at the same time we let $\kappa_i$ tend to zero, in such a way that $\alpha_\tau=\vec\theta_\tau\vec\kappa$ and $\vec\alpha=\vec\theta\times\vec\kappa$ remain constant. We multiply (\ref{kappa}) by $\kappa_i$ and proceed to these limits, as follows.

If $n=0$
\begin{equation}
\kappa^2\simeq-\frac{\lambda \vartheta}{2}\sum_l\int_\Lambda\frac{d^3r}{(2\pi)^3}\frac{\left(2\pi l\vartheta\alpha_\tau +\vec\alpha\vec r\right)\sin\left(2\pi\vartheta l\alpha_\tau+\vec\alpha\vec r\right)}{{r}^2+\hat\mu^2\left[1+\cos\left(2\pi\vartheta l\alpha_\tau+\vec\alpha\vec r\right)\right]}+{\cal O}[(\lambda \vartheta)^2]\label{kappa2}.
\end{equation}

If $n\neq0$
\begin{equation}
\kappa^2\simeq-\frac{\lambda \vartheta}{2}\sum_l\int_\Lambda\frac{d^3r}{(2\pi)^3}\frac{\left(2\pi l\vartheta\alpha_\tau +\vec\alpha\vec r\right)\sin\left(2\pi n\vartheta \vec\theta_\tau\vec r\right)}{{r}^2+\hat\mu^2\left[1+\cos\left(2\pi n\vartheta \vec\theta_\tau\vec r\right)\right]}+{\cal O}[(\lambda \vartheta)^2]\label{kappa3}.
\end{equation}
Both integrals differ by the argument of the trigonometric functions, which is finite when $n=0$, and tends to infinity when $n\neq0$. In both cases the denominator is positive. For $n=0$ it is easy to see that in general  the integral does not vanish. For $n\neq 0$ the trigonometric functions are highly oscillating and could lead the integral to zero. However these oscillations appear in the numerator and in the denominator and have the same frequency, in such a way that the integral does not vanish, as can be numerically verified. Therefore in the limit $\vec\theta_\tau,\vec\theta\rightarrow\infty$, $\vec\kappa$ does not vanish, consistently with \cite{gubser,chen}. 

Summarizing, the Raleigh-Ritz variation of the effective action (\ref{cjt3}) leads to the equations (\ref{kappa}), whose solutions for $\kappa_i$ are different from zero in the same case analyzed in \cite{gubser}, i.e. in the planar limit, when $\Theta_{\mu\nu}\Lambda^2$ tend to infinity, and additionally when $n\neq0$ and with anisotropic non commutativity, i.e. non vanishing $(\Theta^2)_{0i}$.

Returning to the phase transition between the non uniform and the uniform phases, i.e.  at $Q_i\rightarrow 0$, let us define $\Delta G(p_k)=G(p_k,0)-G_0(p_k)$. Thus, from (\ref{g0}) and (\ref{gq0}), it can be shown that
\begin{equation}
\Delta G(p_k)=-\frac{3}{2}\frac{\mu^2}{(p_k^2+2\mu^2)(p_k^2+\frac{7}{2}\mu^2)}+{\cal O}(\lambda).
\end{equation}
Then from this result, together with (\ref{a0}) and (\ref{aq0}), we obtain 
\begin{eqnarray}
\frac{\lambda}{8}[A^2(0)-A_0^2]&=&\frac{1}{4}\mu^2-\frac{3\lambda T}{4}\sum_l\int_\Lambda \frac{d^3q}{(2\pi)^3}\left[\frac{1}{4}G_0(q_l)+\Delta G(q_l)\right]\nonumber\\
&=&\frac{1}{4}\mu^2-\frac{3\lambda T}{16}\sum_l\int_\Lambda \frac{d^3q}{(2\pi)^3}\frac{q_l^2+8\mu^2}{(q_l^2+2\mu^2)(q_l^2+\frac{7}{2}\mu^2)}+{\cal O}(\lambda^2)\nonumber\\
&=&\frac{1}{4} \mu^2-\frac{3\pi\lambda T}{16}
\sum_l\Bigg\{1-4\sqrt{4 l^2 \pi ^2 T^2+2\mu ^2} \arctan\left[\frac{1}{\sqrt{4 l^2 \pi ^2 T^2+2\mu ^2}}\right]+\nonumber\\
&&+3 \sqrt{4 l^2 \pi ^2 T^2+\frac{7 \mu ^2}{2}} \arctan\left[\frac{1}{\sqrt{4 l^2 \pi ^2 T^2+\frac{7 \mu ^2}{2}}}\right]\Bigg\}+{\cal O}(\lambda^2)
\label{aq00},
\end{eqnarray}
An evaluation of the the r.h.s. of this expression can be performed, and it turns out that it is positive, and does not vanish, for all temperatures below the cutoff, and for a perturbative coupling constant. In fact a numerical evaluation shows that the term into the sum decreases quite as $l^{-2}$, and the sum can be performed. Therefore there is a first order phase transition from the uniform to the non-uniform phase. As observed above, this result remains valid for the usual, temperature independent, stripe background.

Equation (\ref{aq00}) can be also applied to the phase transition from the non uniform phase to the disordered phase, with $\phi=0$, hence constant. The r.h.s. of this equation decreases nearby linearly when the temperature increases, but it reaches the maximum temperature (given by the cut-off) before reaching the zero value, so that the corresponding phase transition is as well first order.

\section{Conclusions}
In this work we study the noncommutative effective action at finite temperature for a real
scalar field theory with $\lambda\Phi^4$ interaction, in 4 dimensions, with noncommutative time, and hence temperature. Consequently we take a nonuniform background which depends on imaginary time and temperature. The noncommutativity of imaginary time hints to the possibility of nonequilibrium processes, for which the CJT 2PI effective action is a suitable formalism.

We solve to first order in $\lambda$ the stationarity equations of the CJT effective action, keeping the general form of the propagator $G(x,y)$, with a stripe type ansatz for the background, which depends on temperature. The CJT effective action shows a discontinuity between the results for uniform and non uniform (stripe) background, which is present also in the stationarity equations. We show that this discontinuity originates first order phase transitions between the corresponding phases, in particular between the ordered uniform and non uniform phases, as conjectured in \cite{gubser}. This result is independent of the value of $n$ in (\ref{stripe}). Further, we look for the existence of non uniform solutions, which turn out, as already known \cite{gubser,chen}, in the planar limit of non commutativity, but also in the case anisotropic non commutativity parameters, with explicit temperature dependence of the background, which had not been considered before.

Nonequilibrium processes at high energies are of fundamental interest for the early universe evolution and in QCD. Noncommutative effects have been studied in many aspects in Cosmology and in General Relativity, see e.g. \cite{nc}, and in general for QCD \cite{calmet}, but not as well at finite temperature.

\vskip 2truecm
\centerline{\bf Acknowledgments}
We thank VIEP-BUAP and CONACYT for its support, and thanks Mario Maya for discussions in the early stages of this work. We are also grateful to Hugo Garc\'\i a-Compe\'an, Alejandro Ayala and David Vergara for useful comments.


\end{document}